\documentclass[prd,aps,twocolumn,nofootinbib]{revtex4}
\usepackage[dvipdfm]{graphicx}
\usepackage{amssymb}
\usepackage{amsmath}
\usepackage{parskip}
\usepackage{epsfig}

\newcommand{\bea}{\begin{eqnarray}}
\newcommand{\eea}{\end{eqnarray}}

\newcommand{\be}{\begin{equation}}
\newcommand{\ee}{\end{equation}}
\newcommand{\beqy}{\begin{eqnarray}}
\newcommand{\eeqy}{\end{eqnarray}}
\newcommand{\p}{\partial}

\newcommand{\mx}{\mbox}
\newcommand{\mt}{\mathtt}

\newcommand{\Ga}{\Gamma}

\newcommand{\de}{\delta}
\newcommand{\De}{\Delta}

\newcommand{\e}{\epsilon}

\newcommand{\Om}{\Omega}
\newcommand{\la}{\lambda}
\newcommand{\La}{\Lambda}
\newcommand{\ka}{\kappa}

\newcommand{\cO}{{\cal O}}

\newcommand{\cN}{{\cal N}}

\newcommand{\ra}{\rightarrow}
\newcommand{\Ra}{\Rightarrow}

\newcommand{\LF}{\left(}
\newcommand{\RF}{\right)}
\newcommand{\LT}{\left[}
\newcommand{\RT}{\right]}
\newcommand{\ie}{{\it i.e.\ }}

\newcommand{\lval}{\lambda_{\mt{val}}}

\begin{document}
IGC-09/1-3
\preprint{IGC-09/1-3}
\title{Inflation with a Negative Cosmological Constant}

\author{Tirthabir Biswas~$^a$}
\author{Anupam Mazumdar~$^{b,~c}$}

\affiliation{
{\it $^a$ Department of Physics,
Institute for Gravitation and the Cosmos,
The Pennsylvania State University,
104 Davey Lab, University Park, PA,16802, U.S.A }
\vskip 1mm
{\it $^b$ Physics Department, Lancaster University, Lancaster, LA1 4YB, UK }
\vskip 1mm
{\it $^c$ Niels Bohr Institute, Copenhagen, Blegdamsvej-17, DK-2100, Denmark}
}

\begin{abstract}
We find a unique way of realizing inflation through cyclic phases in an universe with negative vacuum energy. According to the second law of thermodynamics entropy monotonically increases from  cycle to cycle, typically by a constant factor. This means that the scale factor at the same energy density in consecutive cycles also increases by a constant factor. If the time period of the oscillations remain approximately constant then this leads to an ``over all'' exponential growth of the scale factor, mimicking inflation. A graceful exit from this inflationary phase is possible as a dynamical scalar field can take us from  the negative to a positive energy vacuum during the last contracting phase.
\end{abstract}
\date{\today}

\maketitle
\section{Introduction}

Inflation is one of the most popular paradigm which explains large scale homogeneity as well as  structures within
our universe by an epoch of accelerated expansion in the early universe. Inflation requires a large positive vacuum
energy density, i.e. it {\it effectively} gives rise to a deSitter (dS) space-time. On the other hand the superstring theory which
is considered to be one of the most fundamental theory
of nature, and makes a robust claim to combine both gauge theory and gravity,
naturally allows anti deSitter (AdS)  vacua, i.e. a negative cosmological constant \cite{GKP}. It is believed
that various non-perturbative and supersymmetry breaking effects would eventually lift the AdS into dS
in such a way that it would be consistent with the current cosmological observations~\cite{kklt}. Nevertheless, realizing such an uplifting mechanism seems extremely non-trivial, and it is perhaps much more likely that most of the string vacua   have ``large'' negative energies. Is there a way to come out of these negative energy vacua and be consistent with the ``Standard Model'' of inflationary and $\La$CDM cosmology?\\

Another aspect of a dS inflation is that it naturally dilutes all matter and therefore a graceful exit of inflation also requires a successful
reheating of the universe with the observed Standard Model degrees of freedom. However, there are only few notable examples
where the identity of the inflaton can be made successfully within the minimal supersymmetric Standard Model or within  models
with modified gravity and Standard Model Higgs,  thereby
ensuring   successful reheating of the Standard Model degrees of freedom~\cite{AEGM,Higgs}.

The aim of the present paper is to realize inflation in an AdS space-time, or more precisely when the
vacuum energy is negative.  At first glance the  idea of realizing inflation with a negative
cosmological constant seems rather paradoxical. However as we shall show here, if we give up the idea that the universe began in a
{\it cold}  state devoid of any thermal entropy, but rather consider the possibility that it begins in a
{\it hot} thermal state consisting of relativistic species and non-relativistic particles, then we have a chance
to realize ``Cycling inflation''.  An advantage in our case is that it lends the possibility of identifying the thermal state with the Hagedorn phase in string theory.

\begin{figure}[htbp]
\begin{center}
\includegraphics[width=0.40\textwidth,angle=0]{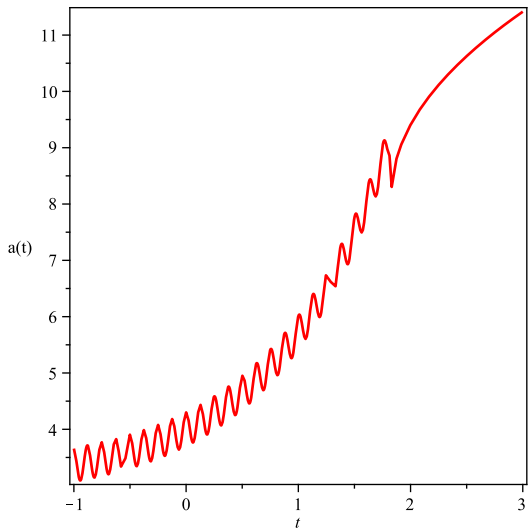}
\end{center}
\caption{Qualitative evolution of the scale factor: After the initial cyclic/inflationary phase, the universe enters an everlasting expanding phase following the last bounce.
  \label{fig:scalefactor}}
\end{figure}

We will obtain inflation in a cyclic universe~\footnote{Cyclic cosmologies have been
 considered in many references, see~\cite{emergent,tolman,narlikar,ekcyclic,barrow,dabrowski,phantom,piao,kanekar,saridakis,columbia}, for
a review see~\cite{review}.} involving non-singular bounces~\footnote{ The readers are referred to several attempts in this
regard  mostly involving  non-local and/or non-perturbative physics, such as string inspired non-local modifications
of gravity~\cite{bms},  stringy toy models using AdS/CFT ideas~\cite{turok}, tachyon dynamics~\cite{tachyon},
mechanisms involving ghost condensation~\cite{ghost}, fermion condensations (both classically~\cite{greene}
and via quantum  BCS-like gap formation~\cite{ab}, brane-world scenarios with extra time-like directions~\cite{sahni} and
in loop quantum cosmology~\cite{loop}.} over a time scale which is much larger compared to the cycle time period. The
cosmological evolution in our model will consist of two distinct phases, see Fig.~[\ref{fig:scalefactor}]:

\begin{itemize}

\item{An inflationary phase where the universe undergoes cycles of expansion and contraction, but it contracts  less than it expands in each cycle in accordance with the second law of thermodynamics. The time period of these cycles are ``short'', approximately constant,
and ``on an average'' the universe seems to be experiencing exponential expansion.}

\item{This above cyclic inflationary phase ends via suitable scalar field dynamics, the universe bounces one last time and then ``exits'' into an everlasting expanding phase resembling our current universe.}

\end{itemize}

In the following section we will discuss how we can obtain the inflationary phase in the presence of a negative cosmological constant.
In section~\ref{exit}, we will elaborate on the last bounce and how one can exit the inflationary phase when one includes the dynamics of a scalar field with an appropriate potential. In section~\ref{conclusion} we will conclude by briefly summarizing the cosmological scenario we have presented, and discuss some of it's advantages as well as open issues that needs to be addressed further. Some computational details are provided in two appendices~\ref{A},\ref{B}.

\section{Cycling and  Inflating}\label{inflation}

In order to realize the first phase, let us assume a simplified  model where the ``matter content''  consists of non-relativistic
and relativistic degrees of freedom and a negative cosmological constant. Let us also assume that both relativistic and
non-relativistic species are in thermal equilibrium above a certain critical temperature, $T_c$, but  below this temperature
the massive non-relativistic degrees of freedom fall out of equilibrium and consequently at later stages
when they decay into radiation  thermal entropy is generated.

The above picture is inspired by the well known stringy Hagedorn phase where all the states, massless and massive, are in thermal equilibrium close to the Hagedorn temperature~\cite{tseytlin,jain,vafa}, but below some critical temperature (not too small compared to the string scale) some of the massive states are expected to fall out of equilibrium. The important implications of the Hagedorn phase in String-Gas-Cosmology~\cite{tseytlin,vafa,string-thermo,anupam,hindmarsh,nayeri,kaya} (see~\cite{string-gas-review} for recent reviews) has been discussed  in the past, and more recently in the context of cyclic models~\cite{emergent,columbia}. In particular, in the cyclic models production of entropy has been found to be a key ingredient in determining the cosmology.

Essentially,  production of thermal entropy  makes the cycles asymmetric; in our simple toy model we will find that in any given cycle the universe expands more than it contracts,  and this is what allows us to mimic inflation. To make the picture clearer, let us call the scale factor at the transition from thermal to non-thermal phase in the $n$th cycle as $a_{c,n}$. Then the average Hubble expansion rate in the $n$th cycle\footnote{We are defining our cycle to ``start'' from the transition point
$a_{c,n}$ and end at the next transition point, $a_{c,n+1}$.} is given by:
\be
<H>\equiv {\int H dt\over \int dt}= {1\over \tau_n}\ln\LF{a_{c,n+1}\over a_{c,n}}\RF\equiv {\cN_n\over \tau_n}\,,
\ee
where $\tau_n$ is the time period of the $n$th cycle. Let us now imagine that the time periods and the average growth factors are approximately constant,  $\tau_n\approx \tau$ and $a_{c,n+1}/ a_{c,n}\approx \exp\cN$, respectively. In this case it is clear that although the universe undergoes oscillations, on an average it maintains a constant Hubble expansion rate. Through the course of many many oscillations the universe can become exponentially large thereby addressing almost all the standard cosmological puzzles (except possibly the monopole problem) in much  the same way as standard inflationary scenarios do. A more detailed discussion on these issues and also on generation of scale-invariant perturbations is presented in the concluding section.

Let us therefore now see how the above ingredients can be achieved in our simple toy model. Our aim is to calculate
$\tau$ and $\cN$, and  check whether they  really remain constants over cycles. Let us divide the dynamics in a given cycle  into a thermal bounce phase and a non-thermal turnaround phase. Without specifying the physics of the bounce, we are here  going to assume that it is  non-singular (for recent progress in this direction the readers are referred to~\cite{emergent,casimir,bms,loop,turok,ghost,tachyon}), and also for simplicity that it occurs
at some fixed energy density, $\rho_b$, in every cycle.

Since during the  bounce phase the universe is assumed to be in thermal equilibrium~\footnote{In a singular big crunch/bang scenario the interaction rates which can potentially maintain thermal equilibrium among the different  species,  cannot possibly keep up with the {\it diverging}  Hubble rate. However in a non-singular bouncing scenario the Hubble rate is bounded from above and therefore it is possible to maintain thermal equilibrium. This point has been discussed in more details in~\cite{emergent} in the context of a Hagedorn physics.}, no entropy is produced and  the evolution is going to be symmetric. Thus, for our purposes the thermal bounce phase is rather uninteresting, there is no overall growth, and we have to look into the non-thermal phase to
realize inflation.

In the non-thermal phase the Hubble equation  reads:
\be
H^2={T_c^4\over 3M_p^2}\LF {\Om_r\over a^4}+{\Om_m\over a^3}-{\La\over T_c^4}\RF\,,
\label{hubble}
\ee
where $-\La$ is the negative cosmological constant, and  the $\Om$'s are related to the energy densities via
\be
\rho_m=T_c^4{\Om_m\over a^3}~~~\mx{ and }~~~\rho_r=T_c^4{\Om_r\over a^4}\,.
\ee
For definiteness, we consider a compact  universe~\footnote{For an open or a flat universe one
just has to rephrase all the arguments in terms of entropy density rather than the total entropy and volume
of the universe.} with a volume $V\equiv T_c^{-3}a^3$. Let us define the ratio of the equilibrium energy densities
of the  non relativistic (massive) and relativistic (massless)   species  at the transition point to be given by
\be
\mu\equiv{\rho_{m,c}\over \rho_{r,c}}\,.
\ee
We are going to be interested in a special case when $\mu\gg1$, and the decay rate of the massive particles to radiation is small  compared to the time period we are interested in.  This means that the non-thermal phase is dominated by matter density, and we can ignore radiation in the Hubble equation. We also want  $T_c^4\gg \La$ to ensure that the turnaround happens in the non-thermal phase once the matter energy density  becomes comparable to $\La$.

In order to understand the dynamics, it is instructive to first look into the evolution when non-relativistic and relativistic
species are non-interacting.  Then $\Omega$'s would just be constants yielding an Friedmann-Robertson-Walker (FRW) cosmology where the universe enters the non-thermal  expansion phase at the transition temperature $T_c$, turns around due to the presence of the negative cosmological constant and then contracts again to the temperature $T_c$, whereafter it enters the thermal (still contracting) phase. If we neglect radiation  in the Hubble equation one can compute the time period of the cycles quite precisely, and it is  given by a rather simple expression~\footnote{We have ignored the time spent in the thermal bounce phase, because approximately it is given by: $\tau_b\sim M_p/T_c^2$, which is much shorter than $\tau$ as long as $T_c^4\gg\La$.} (see the appendix \ref{A} for details):
\be
\tau\approx {3.6M_p\over \sqrt{\La}}\,.
\ee
This is a constant and does not change from cycle to cycle as it does not depend on $\Om_r,\Om_m$ ($\Om$'s will increase from cycle to cycle with the production of entropy). This is crucial to
realizing our scenario and only works because the turnaround is provided by the negative cosmological constant.
If for instance, the turnaround is due to spatial curvature, the time period keeps increasing. Such an ``emergent cyclic phase'' ~\cite{emergent} can in fact precede the inflationary phase  providing geodesic
completeness to our inflationary model, but we  certainly need the negative cosmological constant to realize the inflationary mechanism.

Let us now turn our attention to the average growth factor, which is best expressed in terms of entropy growth from
cycle to cycle. The usual thermodynamic entropies associated with matter and radiation are given by
\be
S_r={4\rho_r V\over 3T}={4\over 3}g^{1/4} \Om_r^{3/4}\,,~~~~S_m={\rho_m V\over M} =\Om_m\,.
\label{entropy-formula}
\ee
Here  $M\approx T_c$ corresponds to the mass of the non-relativistic particles, and in our convention,  $g=(\pi^2/30) g_{\ast}$, where $g_{\ast}$ is the number of ``effective'' massless degrees of freedom. From the above equations we find that at the transition point the ratio of the two entropies are just given by:
\be
{S_m\over S_r}={3\mu \over 4}\,.
\ee
Thus one can express all the quantities in terms of the total entropy, $S=S_m+S_r$, of the system
\begin{eqnarray}
\Om_m={3\mu \over4+3\mu}S\ ; \ \Om_r=\LT{3\over g^{1/4}(4+3\mu)}\RT^{4/3}S^{4/3}\\
\mx{ and }a_c=\LT{3\over (4+3\mu)g^4}\RT^{1/3}S^{1/3}\,,
\end{eqnarray}
where to arrive at the last relation we have used the thermodynamic relation
$\rho_r= gT^4$. This illustrates how all the different quantities grow with the increase in entropy from
cycle to cycle. In particular if we can find by how much the entropy increases in a single cycle we will know by
how much $ a_c$ increases.

Now below $T_c$, entropy is generated via energy exchange between non-relativistic matter and radiation~\footnote{
This is also realizable in the context of a Hagedorn phase as discussed in Ref.~\cite{emergent,columbia}.}.
Phenomenologically, such energy exchanges can be captured  by generalizing  conservation equations~\cite{barrow,dabrowski,emergent} for the two fluids to:
\be
\dot{\rho_r}+4H\rho_r=T_c^4 s\,,~~~~~~~~\dot{\rho_m}+3H\rho_m=-T_c^4s
\label{continuity}
\ee
which now includes an energy exchange term. We can easily compute the net entropy increase:
\be
\dot{S}=\dot{S_r}+\dot{S_m}=a^3s\LF{3b_rT_c\over 4\rho_r^{1/4}}-{T_c\over M}\RF=a^3s\LF{T_c\over T_r}-1\RF
\label{entropy-growth}
\ee
Since, we want to consider matter converting into radiation, we will assume $s>0$. Consistency with $2^{nd}$ law of thermodynamics then means that the quantity within brackets must be positive. This is nothing but the condition that the temperature of the non-relativistic species be greater than that of the relativistic species, so that energy flows from the hotter non-relativistic species to colder radiation in accordance with the $1^{st}$ law of thermodynamics.

Since in our picture the two species have the same temperature $T_c$, at the transition point, where after $T_r$ decreases, while the matter ``temperature'' $T_m=M$ stays fixed, Eq.~(\ref{entropy-growth}) is consistent with both the $1^{st}$ and  $2^{nd}$ law of thermodynamics. Further note that the modified continuity equations (\ref{continuity}), obviously satisfies conservation of the total stress energy tensor. However, the energy exchange term breaks the time-reversal symmetry  which is ultimately responsible for providing the arrow of time in the direction of an increasing entropy.

The energy-exchange function, $s$, typically depends on the different energy densities, Hubble rate and the scale factor and takes different forms depending upon the different processes that one is interested in~\cite{barrow-exchange}. Here we will consider the standard massive particle decay process into radiation, where, $s$ is just given by
\be
s={\Ga \rho_m\over T_c^4}\,,
\ee
$\Ga$ being the decay rate. In this case, the matter continuity equation can be trivially integrated to give
\be
\rho_m=\rho_{m,c}\LF{a_c\over a}\RF^{3}e^{-\Ga t} \ .
\ee
Before proceeding further we are going to specialize to the case when  the decay time is much
larger than the time period of the cycle, or technically, $\Ga^{-1}\gg  M_p/\sqrt{\La}\sim \tau$. The entropy generation
formula then simplifies to:
\be
\dot{S}=\frac{\Ga\rho_{m,c}a_c^3e^{-\Ga t}}{T_c^4}
\LF\frac{a}{a_c}-1\RF=
\frac{3\mu  S_c\Ga e^{-\Ga t}  }{4+3\mu}\LF\frac{a}{a_c}-1\RF\,,
\label{Sdot}
\ee
where by $S_c$, we denote the entropy at the beginning of the cycle at the transition temperature $T_c$. Since we are considering $\Ga^{-1}\gg \tau$, only small amounts of entropy are produced and to obtain a leading order estimate, $\sim\cO(\Ga \tau)$, of the entropy production  we can treat $\Om_m$ to be a constant in the Hubble equation, Eq.~(\ref{hubble}). Using the approximations $\mu,T_c^4/\La\gg 1$, one can integrate the above equation to obtain a rather simple result for the entropy growth (see appendix \ref{A} for details)
\be
\De S\approx 0.71 S_c(\Ga \tau)\LF{\mu g  T_c^4\over \La}\RF^{1/3}\,.
\ee
Thus, we approximately have
\be
\frac{S_{n+1}}{S_n}= 1+\ka\,,~\mx{ where }~\ka\equiv0.71\times(\Ga \tau)\LF\frac{\mu g  T_c^4}{\La}\RF^{1/3}\,,
\ee
 and $ \cN= {\ka/3}$. Therefore, if $\Ga \tau$ is sufficiently small,  the entropy  only increases by a small  factor which does not vary as we go from cycle to cycle, and this is exactly what we wanted.

\begin{figure}[htbp]
\begin{center}
\includegraphics[width=0.40\textwidth,angle=0]{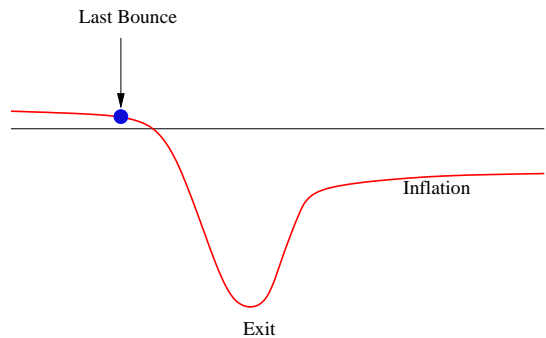}
\end{center}
\caption{Shape of a typical potential.
  \label{pot}}
\end{figure}

\section{Graceful Exit}\label{exit}

If we are stuck with a negative cosmological constant, then the above inflationary phase persists forever and one can never obtain an universe like ours. However, one can exit the inflationary phase if instead of a negative cosmological constant we have a dynamical scalar field whose potential {\it interpolates} between a negative and a positive cosmological constant asymptotically, see Fig.~\ref{pot}. Since $V(\phi)\ra -\La$ as $\phi\ra \infty$, we can realize the inflationary phase, but the scalar field keeps rolling towards smaller $\phi$ and eventually there comes a (last) cycle when in this last contraction phase the scalar field can gain enough energy to zoom through the minimum and reach the dS (positive vacuum energy) phase. In the context of a multi-dimensional string landscape~\cite{landscape} one can think of the inflationary phase as scanning of the minima's with negative energies before finally transiting to positive energies and eventually reaching the vacuum where we are sitting right now, namely the Standard Model or minimal supersymmetric Standard Model vacuum~\cite{Frey}.

\begin{figure}[htbp]
\begin{center}\label{variants}
\includegraphics[width=0.30\textwidth,angle=0]{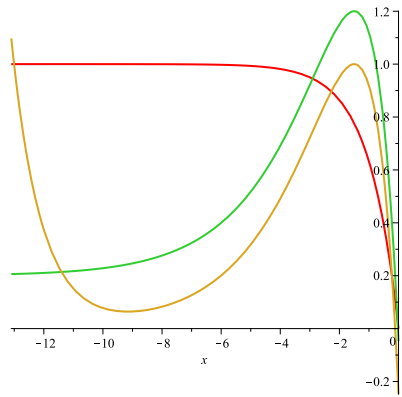}
\end{center}
\caption{Shape of different potentials in the  $\phi<0$ region (dS part of the potential) which can realize a graceful
exit from Cycling inflation. The red curve corresponds to the potential in Fig.~\ref{pot}. The green and the sap green curves respectively correspond to alternatives where either one approaches $V(\phi)\ra 0$ as $\phi\ra -\infty$ like in usual quintessence scenarios, or has a minimum with a small $\sim (\rm mev)^4$ positive cosmological constant.}
\end{figure}

To see this in greater details, let us consider a  potential of a scalar field $\phi$ with the property that it has a minimum at $\phi=0$ and $V(\phi)\ra -\La$ as $\phi\ra \infty$, while $V(\phi)\ra \La_0\sim (\rm mev)^4$ as $\phi\ra -\infty$. For simplicity we will further assume that the potential  increases monotonically  on either side of the minimum~\footnote{If we want to mimic the stringy landscape then we should perhaps add many ripples to the flat parts of the potential. As long as these ripples are small the dynamics should qualitatively be the same as  we discuss here, but it would be interesting to study them in future. Also, the overall shape of the potential that is proposed here is not the only one which will work. Slight variations, as depicted in Fig.~\ref{variants}, should also be fine.}. Our assumptions also mean that $V'(\phi)\ra 0$ as  $\phi\ra \infty$,  the slope, $V'(\phi)$, increases as $\phi$ decreases, reaches a maximum at some point $\phi>0$, and then decreases to zero at the minimum, $\phi=0$.

Now, during the inflationary phase which occurs in the flat negative plateau region, $\phi$ will slowly roll down, it will pick up kinetic energy during contractions and slow down during expansions. The important point to note here is that the amount by which the kinetic energy increases during contraction depends on the slope of the potential. We also know that in a contracting phase with the increase in kinetic energy, the total energy density also increases. This can be seen by simply rewriting the Klein Gordon equation
\begin{eqnarray}
\ddot{\phi}+3H\dot{\phi}=-V'(\phi)\equiv-\la^3 \,,\\
\Ra \dot \rho_{\phi}=-3H\dot{\phi}^2>0\mx{ as long a }H<0\ .
\end{eqnarray}
We can  try to estimate by how much the total energy density increases in a single contraction phase for a constant slope, $\la^3\equiv V'(\phi)$.

In the constant slope approximation  we find (see appendix~\ref{B} for details)
\be
\De \rho_{\phi}\gtrsim {6\la^6M_p^2\over 5\La}\LF{\mu g  T_c^4\over \La}\RF^{5/3}\ .
\ee
At the start of the turnaround the scalar energy density is $\sim -\La$, and therefore in order for the total energy density to become positive we require $\De \rho_{\phi}>\La$ which gives us a rather mild constraint on $\la$:
\be
\la\gtrsim \LF{\La \over M_p}\RF^{1/3}\LF{\La\over \mu g  T_c^4}\RF^{5/18}\equiv \la_{\mt{cr}}
\ee
Thus we note, that if there is a sudden transition from the nearly flat negative scalar potential with slope $\la_{\mt{inf}}\ll \la_{\mt{cr}}$, to a steeply falling valley region with a slope that is large enough, $\la_{\mt{val}}> \la_{\mt{cr}}$, the kinetic energy can overcome the negative potential energy, so that the total scalar energy becomes positive.

Once the total scalar energy density is positive, since the energy density can only increase in the contracting phase, the scalar field cannot turn back in the negative potential region (at the turning point this would imply negative total energy). This is a well known result, see for instance~\cite{linde}. In the absence of a bounce, the scalar field will continue to evolve towards left, enter the positive flat part of the potential, and will become completely dominated by it's kinetic energy, blue shifting as $a^{-6}$. Eventually $\phi$ will go all the way out to $-\infty$ leading to the big crunch singularity at $t=0$.

However, in our scenario before the singularity is reached the universe will  bounce one last time when  the energy density reaches
$\sim\rho_b$. As long as the bounce occurs when  the scalar field is already in, or  is ``sufficiently near'' to,  the flat positive part of the potential, the present universe  will emerge dynamically with a positive cosmological constant after the graceful exit from the Cycling inflationary  phase. Note,  that the universe cannot turnaround any more as the the scalar energy density is no longer negative. Moreover, after the bounce since the kinetic energy of scalar starts to redshift as  $a^{-6}$, even if it dominates matter/radiation, it will  quickly become subdominant ensuring  the entry into a matter/radiation dominated universe. We do not need any additional reheating mechanism. We are assuming that the matter/radiation phase in our model contains SM particles.

Finally, one may be worried that since  the positive potential region has an upward slope, after the bounce the scalar field may be able to turn around and fall into the negative potential region spoiling the standard cosmological ``$\La$CDM'' cosmological scenario. The reason why we should be able to avoid this problem quite easily is because the energy density in
the positive flat part is very small $\sim(\rm mev)^4$, and therefore so should be the slope.
Therefore, once the scalar crosses over to the positive part of the potential, it will essentially not see the potential at all. Initially the dynamics will be completely dominated by it's kinetic energy, which will redshift  as $a^{-6}$, as in the free case.
This will continue till the kinetic energy becomes comparable to the potential energy, after which $\phi$
will indeed turnaround, but it will essentially be completely Hubble damped till  matter density also catches up. At this point
 we will enter the ``current'' acceleration phase, thus reproducing approximately the standard ``$\La$CDM'' cosmology. This part of the evolution is similar to the ekpyrotic scenario~\cite{ekcyclic}.

What happens after the ``current'' dark energy phase? Actually, we have not been completely honest when we claimed that in our model there is a ``Last bounce'' after which the universe enters an everlasting expansion phase.  For a potential such as Fig.~\ref{pot} after $\phi$   turns around, it will again inevitably enter the negative potential region which therefore causes the universe to turn around again!
The time scale of such a turnaround is obviously going to be much larger, at least comparable to our current Hubble rate, so there is no conflict with current cosmological observations. Once the universe  starts to contract its energy density increases, it can go past the minimum again, now moving to the right. At some point the universe will bounce again. If by this time the scalar field has climbed up the negative flat plateau, the  whole ``Big cycle'' can begin all over again. Thus in this context when we say that after the exit from Cycling inflation we have the last expanding phase, we really mean the last in a given  Big cycle. We note in passing that if one wishes to avoid  these interesting complications, one can consider slightly different potentials such as the ones depicted in Fig.~\ref{variants}.

It is clear that the cosmology with the kind of potentials that we have been discussing has rich possibilities and needs a more complete treatment. It is also clear, that apart from  $\la_{\mt{inf}},\  \la_{\mt{val}}$, the success of the exit mechanism (for instance, whether the bounce happens when the scalar field has already reached the positive potential region) will  depend on $\rho_b$, and other parameters governing the potential, such as the mass and depth of the minimum, how sudden the transition is from the plateau to the fall region, etc.  A detailed numerical exploration of the parameter space  will be provided in the future.

\section{Conclusion}
\label{conclusion}

We have presented a unique way of realizing inflation with a negative cosmological constant in a cyclic universe. Although in each cycle the universe expands only a little bit, one can obtain a large number of total inflationary e-foldings, $\cN_{\mt{tot}}$ over many  many cycles.  As long as $\cN_{\mt{tot}}\approx 60$ it is clear that we can explain the present causally connected large and  flat universe, however the issues related to isotropy and homogeneity are more subtle. For instance, one may worry that during the bounce when anisotropies grow it may be lead to a Mixmaster type chaotic behavior.
However,  the scale factor at the bounce point grows as $S^{1/3}$, as the bounce occurs at the same energy density $\rho_b$ in each cycle. Since anisotropy decays as $\sim a^{-6}$ (see for instance a discussion in~\cite{khoury}), it is clear then that once the cyclic behavior starts we are going to be safe from any chaotic Mixmaster evolution in subsequent bounces.

It requires a much more systematic and involved analysis to check  that the  inhomogeneities don't become non-linear and spoil the homogeneity of the universe, and we are presently studying these issues in details~\cite{future}. However, one can make some general arguments which suggest that we may be safe from such problems in our model. In GR the matter fluctuations, $\de\equiv \de\rho/\rho$ can only grow as long as their wavelengths are larger than the Jean's length, $\la_J$,  given by
\be
\la_J\sim c_s{M_p\over\rho}\mx{ where } c_s^2\equiv {\p p\over \p \rho}
\ee
is the sound velocity square. Now, in our scenario even in the matter dominated phase we have some amount of radiation. In the matter phase the Jeans length is shortest\footnote{Although the abundance of radiation (and with it the sound velocity) decreases as the universe expands, the energy density decreases even faster,  the net result being an increase in the Jean's length.} precisely at the transition from radiation to matter,
\be
\la_{J\min}\sim {M_p\over \mu T_c^2}
\ee
Thus during matter phase only perturbations with physical wavelengths larger than $\la_{J\min}$ can grow. On the other hand, once the wavelengths become large enough that their wavelengths become larger than the cosmological time scale, $\la_{\mt{cos}}\sim M_p/\sqrt{\La}$, they become super-Hubble fluctuations and evolves according to the Poisson equation:
\be
\de_k= {k^2\Phi_k\over a^2\rho}
\ee
where $\Phi_k$ is the Newtonian potential characterizing the metric perturbations. Now, in the super-Hubble phase $\Phi_k$ becomes a constant\footnote{There is actually a mode of $\Phi_k$ that grows during contraction, but the same mode starts to decay during expansion after the bounce~\cite{matching}, and since in our scenario the universe expands more than it contracts, we should have a net decaying mode. There may also be some corrections coming from new physics which resolves the singularity at the bounce, but for a non-singular bouncing solution, we do not expect any drastic behavior for the perturbations. Moreover, the bounce time scale is much shorter than the time period of the cycle. Again, this is something we are trying to study currently in more details.}  while $\rho$ oscillates between a minimum and a maximum energy density. Thus we have
$$\de_k< {k^2\Phi_k\over a^2\rho_{\min}}\sim  {k^2\Phi_kM_p\over a^2\sqrt{\La}}$$
In other words $\de_k$ falls as $a^{-2}$ in the super Hubble phase just as in ordinary inflation. Physically what is happening is that although the matter fluctuations can grow, they are getting diluted over the course of many many cycles. The energy density on the other hand is getting created at every cycle. The overall effect is that $\de\equiv\de\rho/\rho$ keeps decreasing. Thus what the above arguments suggest is that fluctuations on a given comoving scale can only grow when their physical wavelengths lie between $\la_{J\min}<\la<\la_{\mt{cos}}$ and therefore can only grow a finite amount. Thus as long as the initial fluctuations are small enough, we should be safe from these inhomogeneities. Whether this constitutes a fine-tuning problem is something that requires a much more detailed investigation with is currently underway~\cite{future}.

Analyzing the detail spectrum of inhomogeneities is however a much more involved task. It may be tempting to argue that if the time period of the cycle is much shorter as compared to the average Hubble expansion rate, then as a first approximation the perturbations may not see the effect of the cycles which would then lead to the usual generation of near scale-invariant perturbations. One can see that since the universe is experiencing an overall inflationary growth, for any given mode there will come a time when the wavelength becomes much larger than the cosmological time scale and therefore freeze. Similarly, going back in the past there comes a time when the wavelength is in the far ultra-violet and hence just oscillate as in the  sub-Hubble phase in standard inflation. The transition from sub to super Hubble phase however is going to be more extended and complex to understand in our scenario and only a detailed analysis can determine whether we can retain approximate scale-invariance of the spectrum.

Finally, we would like to end our discussion by mentioning that we have provided an unique example of realizing inflation
through cyclic phases in an anti deSitter universe in a regime where massless and massive non-relativistic degrees of freedom are also interacting. Inflation ends via a transition from a negative cosmological constant to a positive cosmological constant and this also marks the graceful exit from the cyclic phase. Our scenario therefore opens up the possibility of scanning the negative potential region in the string landscape before finally making the transition to the universe as we observe it now.

As emphasized before, whether one can generate a near scale-invariant spectrum of fluctuations in our model remains a crucial open issue, but at the very least the Cycling inflationary phase may be able to provide  ideal initial conditions for a subsequent low scale inflation.


\section{Appendix A: $\tau$ and $\De S$}\label{A}

Here we calculate the approximate time period and the increase in entropy in a given cycle. We will calculate this under the approximation that radiation can be neglected as $\mu\gg 1$ and treat $\Om$ as a constant as the amount of matter decay in a given cycle is negligible.

To calculate the time period we start by rewriting the Hubble equation:
\be
dt={da\over\dot{a}} = {\sqrt{3}M_p da\over T_c^2a\sqrt{{\Om_m\over a^3}-{\La\over T_c^4}}}
\ee
so that the time period is given by
\be
\tau\approx {2\sqrt{3}M_p \over T_c^2}\int_{a_c}^{a_T} {da \over a\sqrt{{\Om_m\over a^3}-{\La\over T_c^4}}}
\ee
where we have neglected the duration in the thermal bounce phase as it is going to be much shorter than the above integral.
Now at the transition point we have
\be
\rho_{m,c}=\mu\rho_{r,c}=\mu g T_c^4\Ra \Om_m =\mu g a_c^3
\ee
and the turnaround scale factor is given by
\be
a_T=\LF{\Om_m T_c^4\over \La}\RF^{1/3}=a_c\LF{\mu g  T_c^4\over \La}\RF^{1/3}\equiv {a_c\over \e}
\ee
Thus the above integral can be re-expressed as
\be
\tau= {2\sqrt{3}M_p \over \sqrt{\La}}\int_{a_c}^{a_T} {da \over a\sqrt{{g\mu T_c^4\over \La}\LF{a_c\over a}\RF^3-1}}= {2\sqrt{3}M_p \over \sqrt{\La}}\int_{\e}^{1} {\sqrt{y}dy \over \sqrt{1-y^3}}
\ee
Since $\e\ll 1$, we have
\be
\tau\approx {2\sqrt{3}M_p \over \sqrt{\La}}\int_{0}^{1} dy\sqrt{y \over 1-y^3}\approx {3.6 M_p \over \sqrt{\La}}
\ee

In a similar manner we can proceed to calculate the approximate entropy increase. Firstly, since we have assumed that $\Ga \tau\ll 1$, we can approximate $e^{-\Ga t}\approx 1$ in the expression for $\dot{S}$. From (\ref{Sdot}) and (\ref{hubble}) we then find
\be
{dS\over da}={\sqrt{3}\Ga S_cM_p\over T_c^2}{\LF{a\over a_c}-1\RF\over a \sqrt{{\Om_m\over a^3}-{\La\over T_c^4}}}
\ee
Thus the increase in entropy in a given cycle can be calculated as
$$
\De S={2\sqrt{3}\Ga S_cM_p\over T_c^2}\int_{a_c}^{a_T} {da\over a}{\LF{a\over a_c}-1\RF\over \sqrt{{\Om_m\over a^3}-{\La\over T_c^4}}}$$
$$={2\sqrt{3}\Ga S_cM_p\over \sqrt{\La}\e }\int_{\e}^{1} {dy\ \sqrt{y}(y-\e) \over \sqrt{1-y^3}}$$
$$\approx {2\sqrt{3}\Ga S_cM_p\over \sqrt{\La}\e }\int_{0}^{1} dy{y^{3/2} \over \sqrt{1-y^3}}
$$
Thus we finally have
\be
\De S\approx {2.6\Ga S_cM_p\over \sqrt{\La}\e }\approx {0.71S_c(\Ga \tau)\over \e }
\ee

\section{Appendix B: Last contraction phase and energy increase}\label{B}
Here we will try to estimate by how much the energy density increases in the last contraction phase when the scalar field is steeply rolling down towards the valley, approximately with constant slope $\lval$. The Klein Gordon equation for the scalar field can be rewritten as
\be
{d(Ka^6)\over da}={\lval^3a^2\over H}\sqrt{Ka^6}\mx{ where }K={\dot{\phi}^2\over 2}
\label{kinetic-eqn}
\ee
Now, $H$ is given by
\be
H=-{\sqrt{\La}\over M_p}\LT\LF{a_T\over a}\RF^3+{\rho_{\phi}\over \La}\RT
\ee
where $\rho_{\phi}$ is the energy density of the scalar field. At the start of the contraction phase $\rho_{\phi}\sim -\La$ and then it increases towards zero. Our main aim is to check whether the energy density can become positive in the contraction phase.
Integrating (\ref{kinetic-eqn}) we  have
$$
\sqrt{K}a^3=\sqrt{K}_Ta_T^3+{\lval^3M_p\over \sqrt{\La}}\int_a^{a_T}{da\ a^2\over \LT\LF{a_T\over a}\RF^3-{\rho_{\phi}\over \La}\RT^{1/2}}
$$
$$>{\lval^3M_p\over \sqrt{\La}}\int_a^{a_T}{da\ a^{7/2}\over a_T^{3/2}}
$$
where the last inequality is valid as long as the scalar energy density is negative, \ie $\rho_{\phi}<0$. We can now perform the integration easily, we have
\be
K>{\lval^6M_p^2\over \La}\LT\LF{a_T\over a}\RF^3-\LF{a\over a_T}\RF^{3/2}\RT^2
\label{K-bound}
\ee

We are now ready to look at the increase in the total energy density. Again we start by rewriting the Hubble equation as
\be
{d\rho_{\phi}\over da}=-{6K\over a}\Ra \De \rho_{\phi}=6\int_{a_c}^{a_T}{da\ K\over a}
\ee
where $\De \rho_{\phi}$ is the amount of energy that can increase in the non-thermal contracting phase. We can use the lower bound on the kinetic energy derived above (\ref{K-bound}) to obtain a lower bound on the energy increase as well:
\bea
\De\rho_{\phi}&>&{6\lval^6M_p^2\over \La}\int_{a_c}^{a_T}{da\over a}\LT\LF{a_T\over a}\RF^3-\LF{a\over a_T}\RF^{3/2}\RT^2\\
&=&{6\lval^6M_p^2\over \La}\int_{\e}^1dx\LT x^{-3}-x^{3/2}\RT^2\\
&=&{3\lval^6M_p^2\over 10\La}\LT4\e^{-5}-5\e^{4}-80\e^{-1/2}+81\RT
\eea
Since $\e\ll 1$ we have
\be
\De \rho_{\phi}\gtrsim {6\lval^6M_p^2\over 5\La\e^{5}}
\ee
At the start of the turnaround the scalar energy density is $-\La$, and therefore in order for the total energy density to become positive we require $\De \rho_{\phi}>\La$ which gives us a rather mild constraint on $\lval$:
\be
\lval\gtrsim {\La^{1/3} \e^{5/6}\over M_p^{1/3}}
\ee

\acknowledgements

 We would like to thank Robert Brandenberger for extensive discussions and helpful suggestions. TB would like to thank Stephon Alexander, Thorston Battefeld, Daniel Chung, Gary Shiu  and   Paul Steinhardt for their suggestions and critical comments. AM is partly supported by  ``UNIVERSENET''(MRTN-CT-2006-035863) and  by STFC  Grant PP/D000394/1. TB would also like to acknowledge the theory group of UMN, and especially Marco Peloso, for their hospitality.


\end{document}